\documentclass{ws-ijmpd}
\usepackage{subfigure}

\begin{document}

\markboth{F.L. Vieyro, G.E. Romero \& G.S. Vila}
{Non-thermal radiation from Cygnus X-1 corona.}

%
\catchline{}{}{}{}{}
%

\title{NON-THERMAL RADIATION FROM CYGNUS X-1 CORONA.}  

\author{Florencia L. Vieyro, Gustavo E. Romero \& Gabriela S. Vila}

\address{Instituto Argentino de Radioastronom\'ia (IAR, CCT La Plata - CONICET),C.C.5, 1894\\
Villa Elisa, Argentina\\
fvieyro@iar-conicet.gov.ar}

\maketitle

\begin{abstract}
Cygnus X-1 was the first X-ray source widely accepted to be a black hole candidate and remains among the most studied astronomical objects in its class. The detection of non-thermal radio, hard X-rays and gamma rays reveals the fact that this kind of objects are capable of accelerating particles up to very high energies. 

In order to explain the electromagnetic emission from Cygnus X-1 in the low-hard state we present a model of a black hole corona with both relativistic lepton and hadron content. We characterize the corona as a two-temperature hot plasma plus a mixed non-thermal population in which energetic particles interact with magnetic, photon and matter fields. Our calculations include the radiation emitted by secondary particles (pions, muons and electron/positron pairs). Finally, we take into account the effects of photon absorption. We compare the results obtained from our model with data of Cygnus X-1 obtained by the COMPTEL instrument.\end{abstract}

\section{Basic scenario}	

\begin{figure}[!h]
\centering
\includegraphics[clip,width=0.5\textwidth, keepaspectratio]{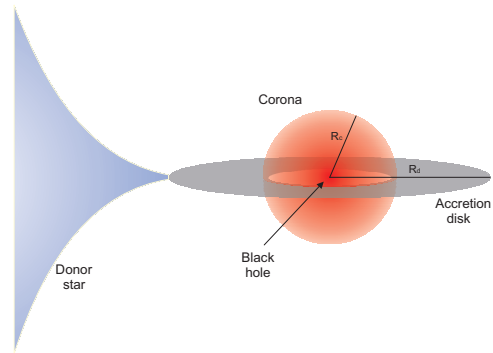}
\caption{Schematic representation of the components of the system. (Not to scale)}
\label{fig:geometria}
\end{figure}

The low-hard state of accreting black holes is characterized by the presence of a hot corona around the compact object. Figure \ref{fig:geometria} shows a scheme of the main components of the system. For this geometry, we assume a spherical corona with a radius $R_{\rm{c}}$ and an accretion disk that penetrates the corona up to a radius $R_{\rm{d}}<R_{\rm{c}}$. We suppose that the corona is homogeneous and in steady state. 

We assume that the luminosity of the corona is 1 \% of the Eddington luminosity of the black hole. For a corona characterized by the parameters of Table \ref{tab_1}, this yields $L_{\rm{c}}=1.3 \times 10^{37}$ erg s$^{-1}$. The corona is composed of a two-temperature plasma, with an electron temperature $T_{e}=10^{9}$ K and an ion temperature $T_{i}=10^{12}$ K (e.g. Ref. \refcite{Narayan}). This is a consequence of electron cooling being more efficient than proton.

In the low-hard state we can assume the system is in a steady state. We then consider equipartition of energy between the different components of the system and obtain the values of the relevant physical parameters. 

The X-rays emission of the corona is characterized by a power law with an exponential cut-off at high energies,
\begin{displaymath}
				n_{\rm{ph}}(E) = \left\{ \begin{array}{lrl}
				A_{\rm{ph}} E^{-\alpha} &  E & < E_{\rm{c}}\\
				A_{\rm{ph}} E^{-\alpha}  e^{-E/E_{\rm{c}}}  &   E & \geq E_{\rm{c}}. 
				\end{array} \right.	
			\end{displaymath}

\noindent In accordance with what is observed in Cygnus X-1 (e.g. Ref. \refcite{Poutanen}), we adopt $\alpha=1.6$ and $E_{\rm{c}}=150$ keV. 
			
Table \ref{tab_1} shows the values of the parameters in our model. 			
			
			\begin{table}[!h]
   \tbl{Model parameters.}
{\begin{tabular}{@{}lc@{}} \toprule

Parameter & Value\\[0.01cm]
\hline\noalign{\smallskip}
$M_{\rm{BH}}$:  black hole mass [$M_{\odot}$]							& $10$$^{(1)}$					\\[0.01cm]
$R_{\rm{c}}$:   corona radius [cm] 										 	  & $5.17 \times 10^{7}$$^{(1,2)}$ 	\\[0.01cm]
$T_{e}$:        electron temperature [K] 									&	$10^9$								\\[0.01cm]
$T_{i}$:        ion temperature [K] 											&	$10^{12}$ 						\\[0.01cm]
$E_{\rm{c}}$:   X-ray spectrum cut-off [keV]							& $150$       					\\[0.01cm]
$\alpha$: 			X-ray spectrum power-law index    				& $1.6$									\\[0.01cm]
$\eta$: 				acceleration efficiency 									& $10^{-2}$							\\[0.01cm]
$B_{\rm{c}}$: 	magnetic field [G]				 								& $2 \times 10^7$				\\[0.01cm]
$n_{i},n{e}$:   plasma density [cm$^{-3}$] 								& $6.2 \times 10^{13}$	\\[0.01cm]
$a$: 						hadron-to-lepton energy ratio 						& $1$ - $100$						\\[0.01cm]
$kT$:						disk characteristic temperature [keV] 	  & $0.1$									\\[0.01cm]
$v$:						advection velocity  [c] 	 		& $0.1$									\\[0.01cm]

\hline\\[0.005cm]
\multicolumn{2}{l}{
$^{(1)}$ Typical value for Cygnus X-1 (Ref. \refcite{Poutanen}).}				  \\[0.01cm]
\multicolumn{2}{l}{
$^{(2)}$ $35 R_{\rm{G}}$.} 			 																				\\[0.01cm]
\end{tabular} \label{tab_1}}
\end{table}

\section{Particle injection}

The injection function for relativistic protons and electrons is a power-law in the energy of the particles $Q(E)=Q_{0} E^{-\Gamma}e^{-E/E_{\rm{max}}}$, which is the consequence of diffusive particle acceleration by shock waves in the coronal plasma. Typical spectral indices are $\Gamma \sim 2.2$ (Ref. \refcite{Drury}). The normalization constant $Q_{0}$ can be obtained from the total power injected in relativistic protons and electrons, $L_{\rm{rel}}=L_{p}+L_{e}$. This power $L_{\rm{rel}}$ was assumed to be a fraction of the luminosity of the corona, $L_{\rm{rel}}=\kappa L_{\rm{c}}$, with $\kappa=10^{-2}$. The way in which energy is divided between hadrons and leptons is unknown, but different scenarios can be taken into account by setting $L_{p}=aL_{e}$. We consider models with an injection dominated by protons in which $a=100$ and models with $a=1$.

\section{Particle losses and acceleration}

The injected particles will lose their energy by different radiative processes. There are three relevant processes of interaction of relativistic electrons with magnetic, matter and photon fields in this scenario: synchrotron radiation, inverse Compton scattering, and relativistic Bremsstrahlung. For protons there are also three relevant processes: synchrotron radiation, proton-proton inelastic collisions and photohadronic interactions. These interactions produce pions, which then decay producing muons. We also consider the energy loss of these particles; the processes for charged pions are the same as for protons and for muons the same as for electrons. 

We consider two mechanisms of particle escape from the corona: advection and diffusion. In the case of advection, particles fall onto the compact object at a mean radial velocity $v = 0.1c$ (Ref. \refcite{Begelman}). Therefore the advection rate is
	
	\begin{equation}
				t_{\rm{adv}}^{-1}=\frac{v}{R_{\rm{c}}} .
		\end{equation}		

\noindent In the case of diffusion, we consider that the corona is static and diffusion of relativistic particles occurs in the Bohm regime. The diffusion coefficient is $D(E)=r_{\rm{g}}c/3$, where $r_{\rm{g}}=E/(eB)$ is the giroradius of the particle. The diffusion rate is

	\begin{equation}
				t_{\rm{diff}}^{-1}=\frac{2D(E)}{R_{\rm{c}}^2} .
		\end{equation}

\begin{figure}[!h]
\centering
\subfigure[Electron losses.]{\label{fig:perdidas:a}\includegraphics[width=0.45\textwidth,keepaspectratio]{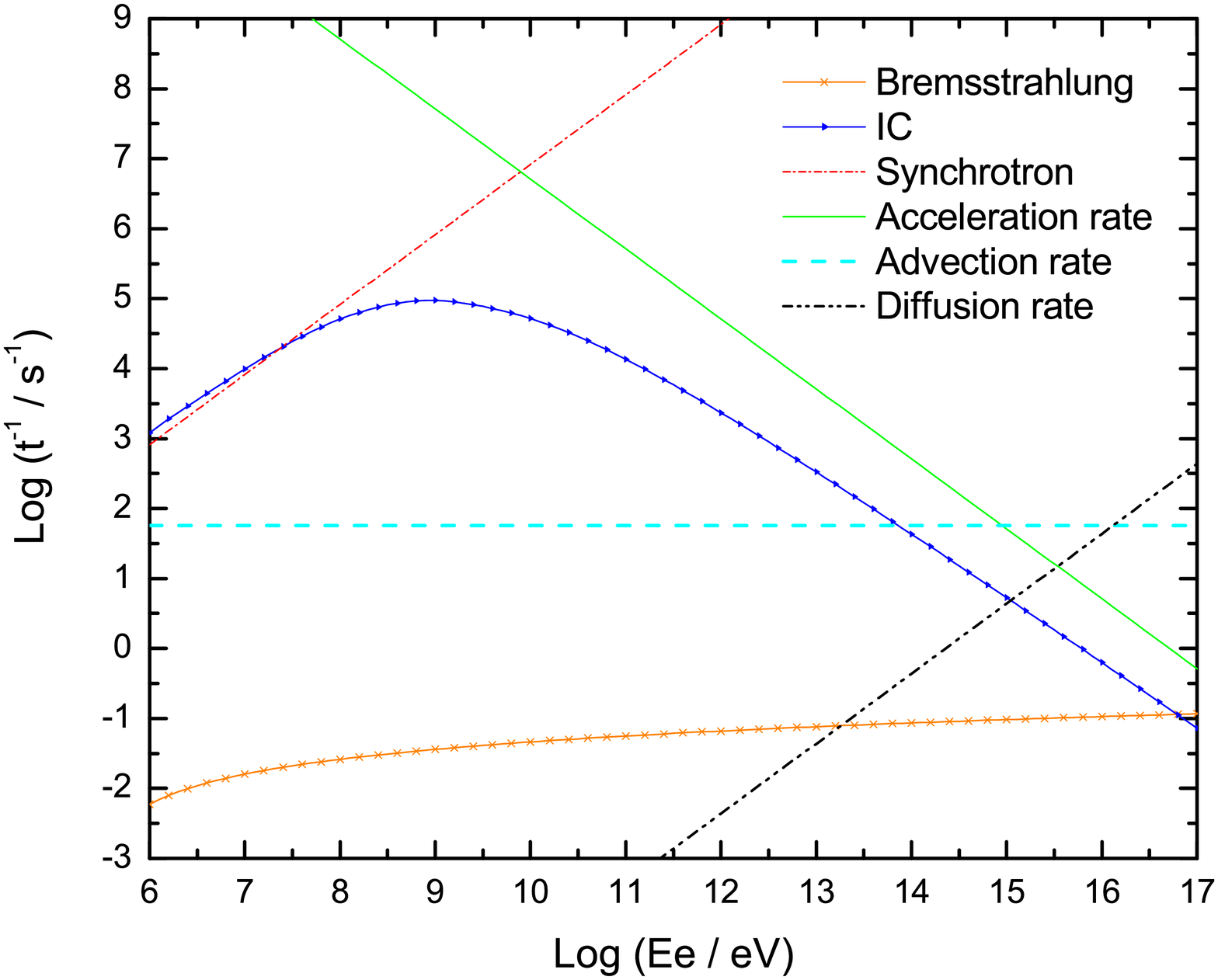}} \hspace{20pt} 
\subfigure[Proton losses.]{\label{fig:perdidas:b}\includegraphics[width=0.45\textwidth,keepaspectratio]{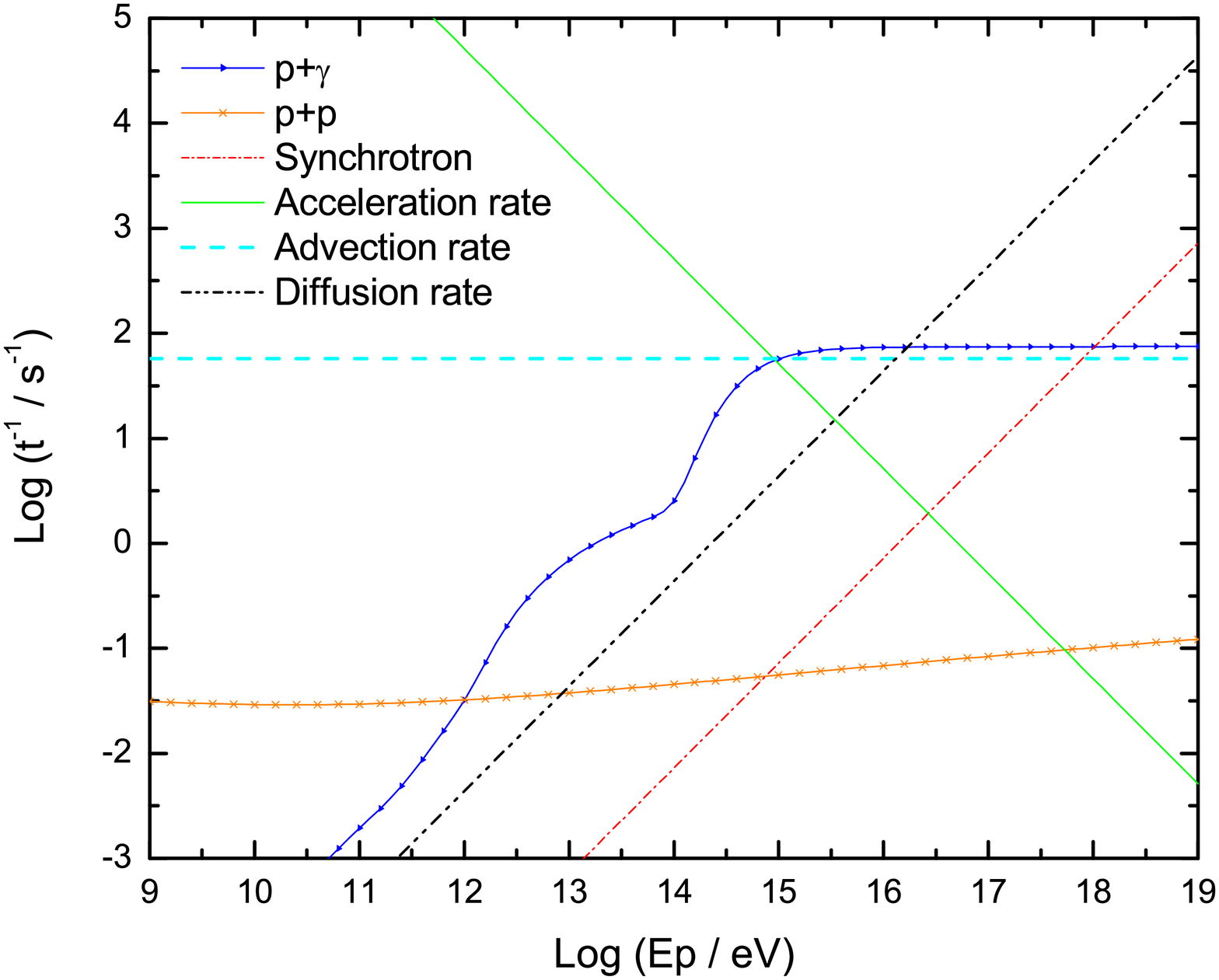}} \hfill \\ 
\subfigure[Pion losses.]{\label{fig:perdidas:c}\includegraphics[width=0.45\textwidth, keepaspectratio]{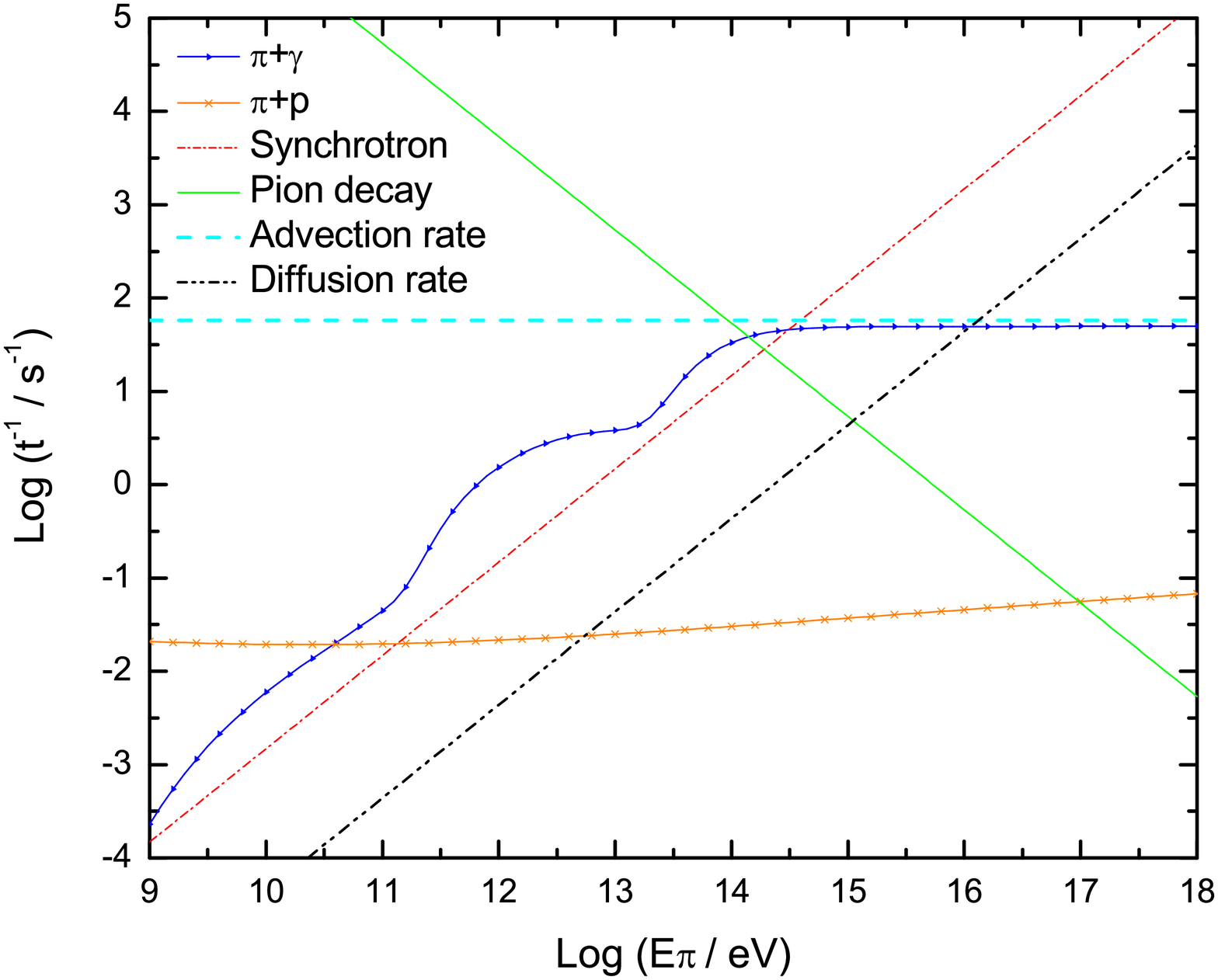}}  \hspace{20pt}
\subfigure[Muon losses.]{\label{fig:perdidas:d}\includegraphics[width=0.45\textwidth, keepaspectratio]{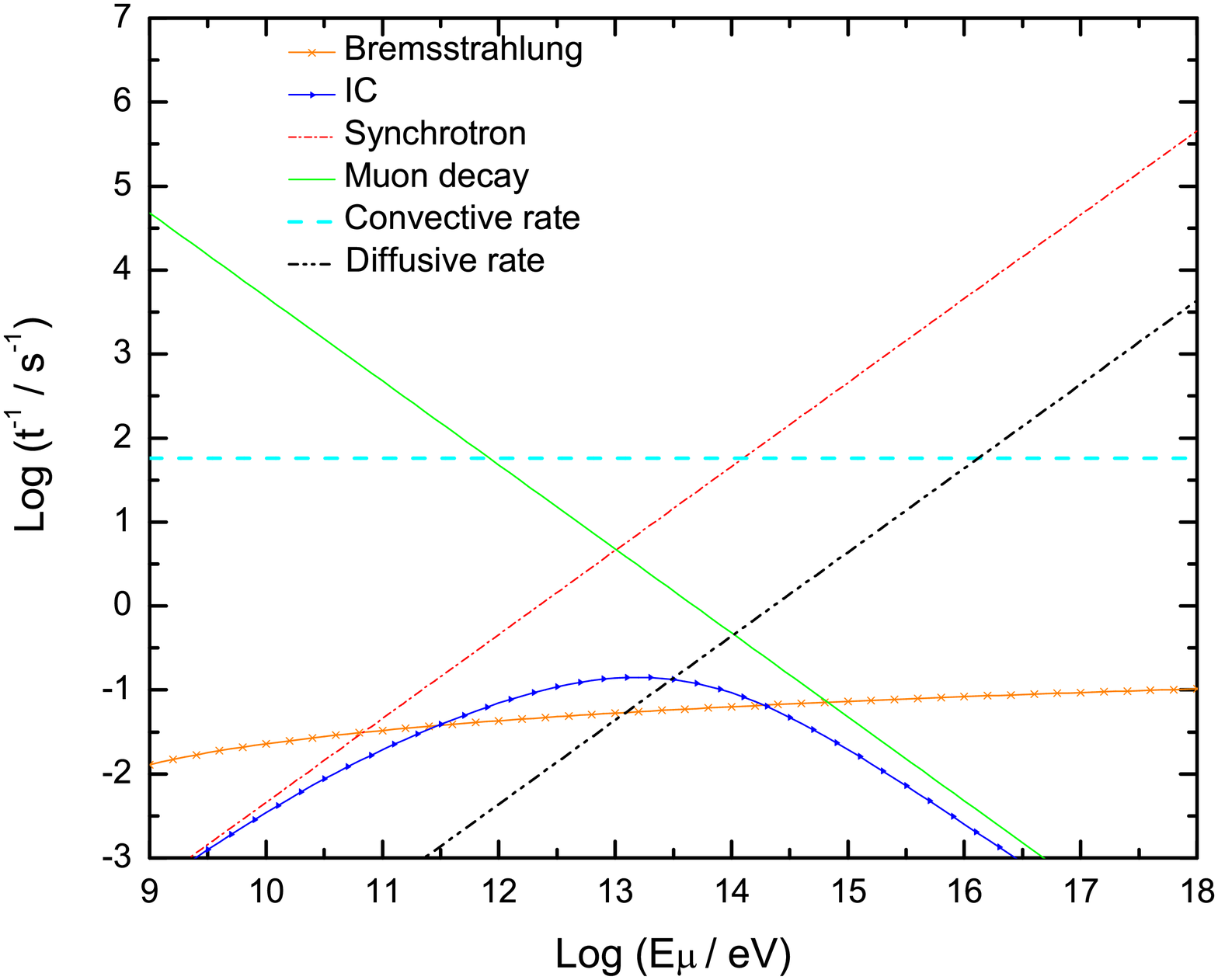}} \hfill
\caption{Radiative losses in a corona characterized by the parameters of Table \ref{tab_1}.}
\label{fig:perdidas}
\end{figure}

	Figure \ref{fig:perdidas} shows the cooling rates for different processes of energy loss, along with the acceleration and escape rates, for each type of particle considered. Under the physical conditions previously described, the main channel of energy loss for electrons is synchrotron radiation. For protons, both $pp$ and $p \gamma$ interactions are relevant. However, in the model with advection, most protons fall into the black hole before radiating their energy.	
	For pions, the main channel of energy loss is $\pi \gamma$ interaction, but an important fraction of pions decay before cooling. On the contrary, muons with energies above $10^{11}$ eV cool mostly by inverse Compton scattering. 
	
The maximum energy that a relativistic particle can attain depends on the acceleration mechanism and the different processes of energy loss. The acceleration rate $t^{-1}_{\rm{acc}}$ for a particle of energy $E$ in a magnetic field $B$ is given by

\begin{equation}
	t^{-1}_{\rm{acc}}=\frac{\eta ecB}{E},
	\label{accrate}
\end{equation}
   
\noindent where $\eta\leq1$ is a parameter that characterizes the efficiency of the acceleration. We have fixed $\eta=10^{-2}$, which describes an efficient acceleration.

\section{Spectral energy distributions (SEDs)}

In order to obtain the spectral energy distributions produced by the different radiative processes, we solve the transport equation for each kind of particle. We calculate the SEDs of different processes following Ref. (\refcite{Vila}) as standard reference. We also calculate the radiation emitted by secondary pairs, which are injected mainly by photon-photon annihilations. 

\begin{figure*}[!t]
\centering
\subfigure[$a=1$, diffusion.]{\label{fig:SEDs:a}\includegraphics[width=0.45\textwidth, keepaspectratio]{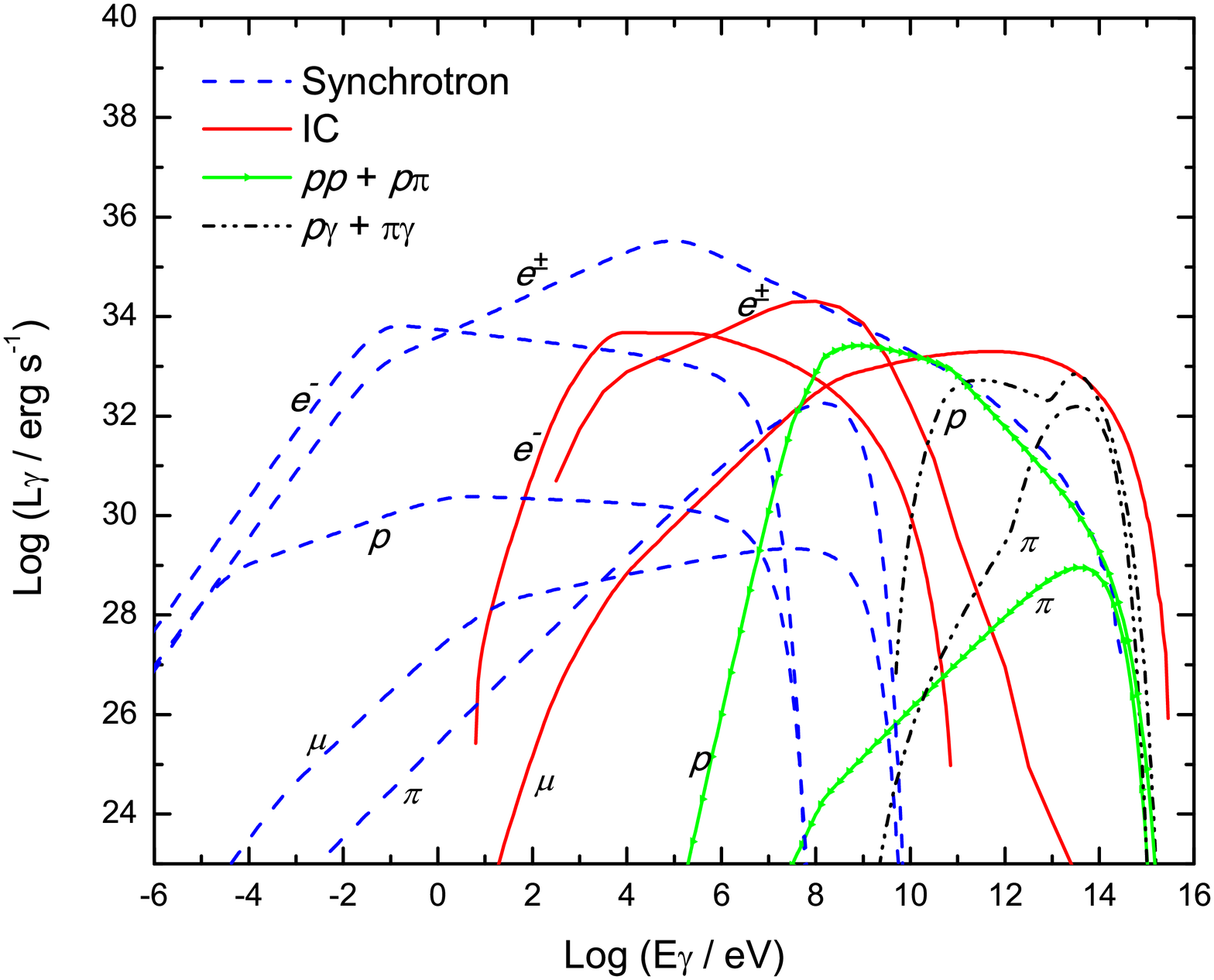}} \hspace{20pt}
\subfigure[$a=1$, advection.]{\label{fig:SEDs:b}\includegraphics[width=0.45\textwidth, keepaspectratio]{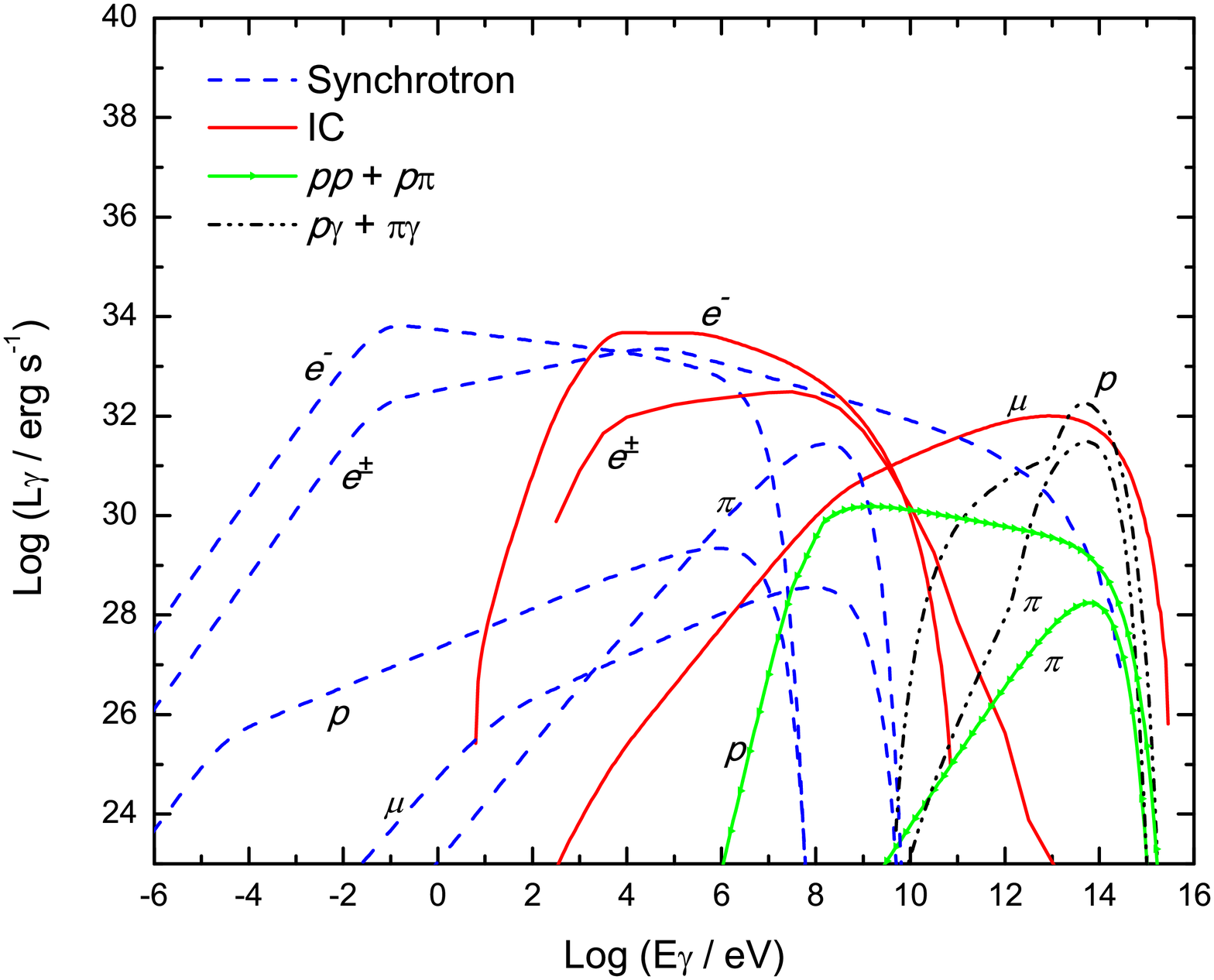}} \hfill \\
\subfigure[$a=100$, diffusion.]{\label{fig:SEDs:c}\includegraphics[width=0.45\textwidth, keepaspectratio]{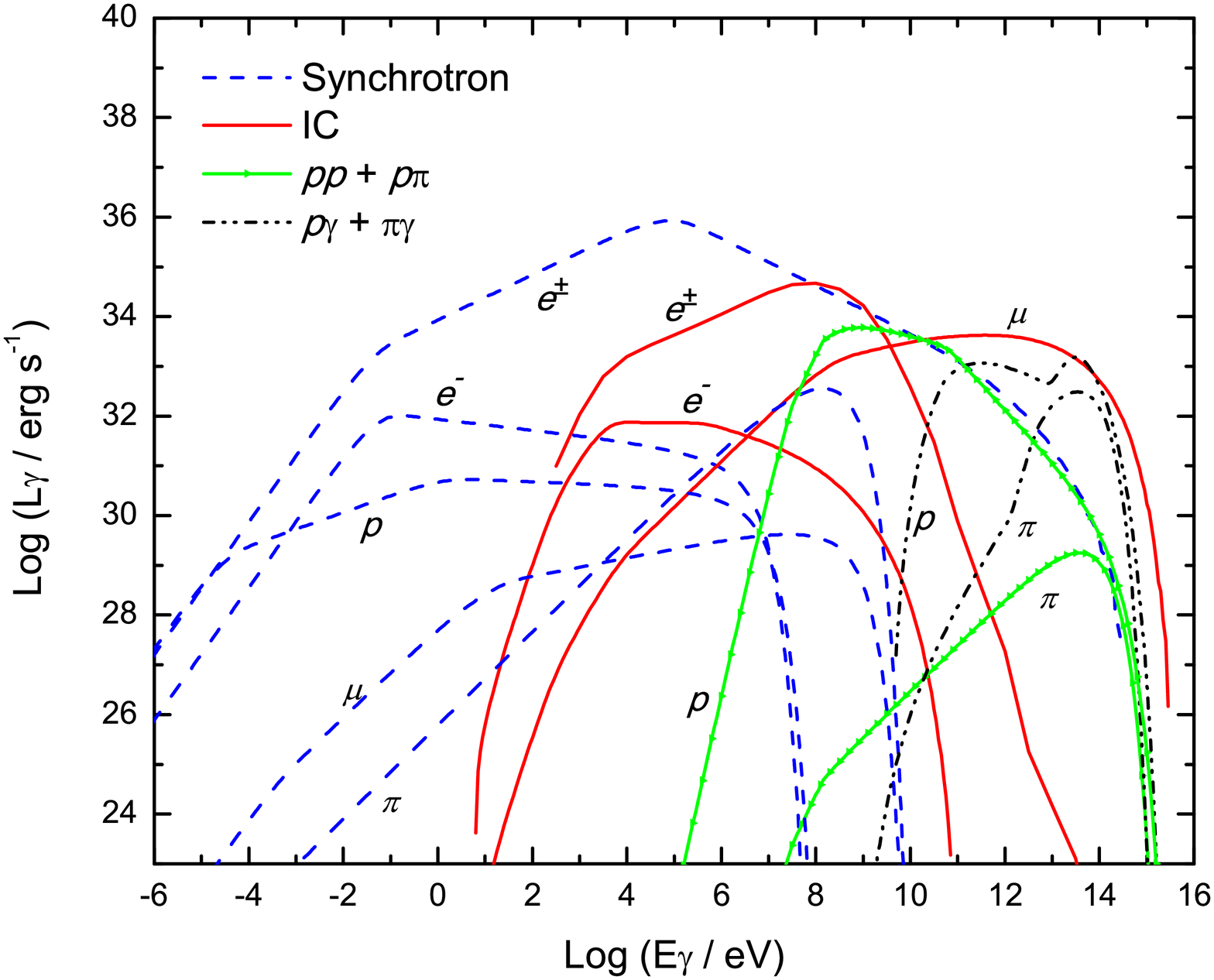}} \hspace{20pt}
\subfigure[$a=100$, advection.]{\label{fig:SEDs:d}\includegraphics[width=0.45\textwidth, keepaspectratio]{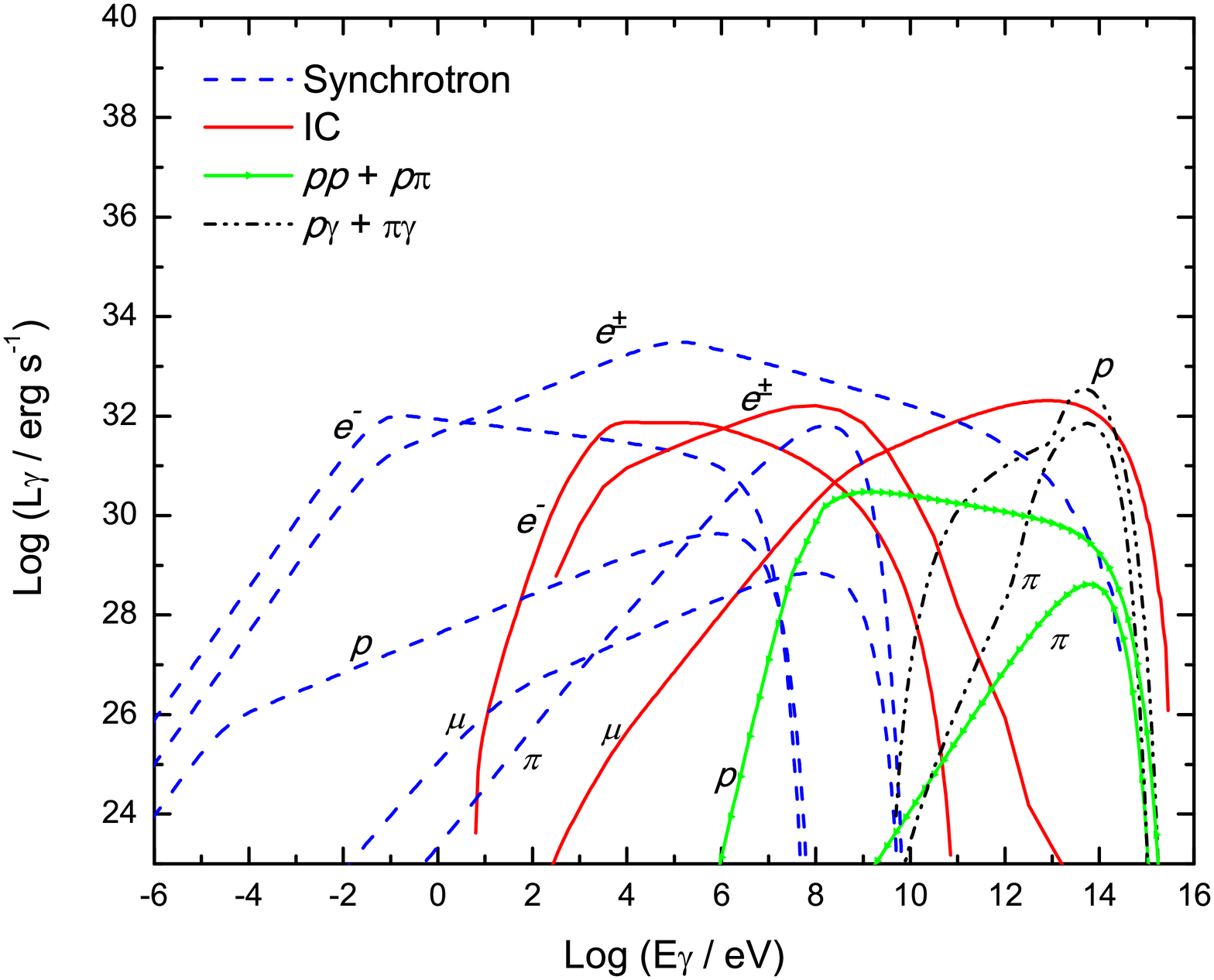}} \hfill 
\caption{Spectral energy distributions obtained for different set of parameters.}
\label{fig:SEDs}
\end{figure*}

Figure \ref{fig:SEDs} shows all contributions to the total luminosity for different advection regimes and for different values of the parameter $a$. Luminosities produced by hadrons and muons are higher in models with diffusion. This is because in models with advection an important fraction of protons and pions are swallowed by the black hole, while with diffusion these particles are able to lose their energy before falling into the compact object or escaping.

Finally, we take into account the effects of photon-photon attenuation. The absorption can be quantified through the opacity $\tau$. If the initial gamma-ray luminosity is $L_{\gamma}^0(E_{\gamma})$, the attenuated luminosity $L_{\gamma}(E_{\gamma})$ after the photon travels a distance $l$ is $L_{\gamma}(E_{\gamma})=L_{\gamma}^0(E_{\gamma}) e^{-\tau(l,E_{\gamma})}$.

\begin{figure}[!t]
\centering
\includegraphics[width=0.7\textwidth, keepaspectratio]{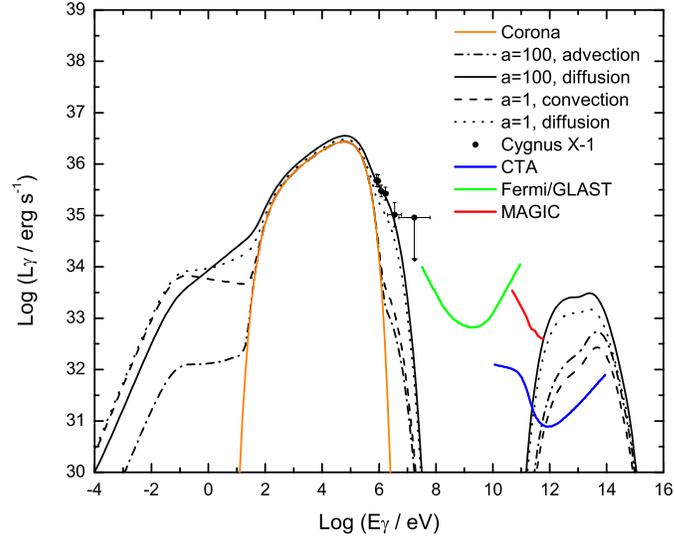}
\caption{Spectral energy distributions. Observational data from COMPTEL of Cygnus X-1.}
\label{fig:TODAS}  
\end{figure}

Figure \ref{fig:TODAS} shows the attenuated luminosity for each set of parameters, the spectrum of Cygnus X-1 as observed by COMPTEL (Ref. \refcite{McConnell}), and the sensitivity of different instruments. Notice that in addition of accounting for the MeV tail observed by COMPTEL, a high energy bump with luminosities $\sim 10^{33}$ erg s$^{-1}$ is expected at $E_{\gamma} > 1$ TeV.

The existence of the high-energy tail in the spectrum detected by COMPTEL and reproduced by our model was recently confirmed by observations with INTEGRAL satellite (Ref. \refcite{Cadolle}).

\section{Conclusions}

On one hand, the SED predicted by the model with an injection dominated by protons and diffusion reproduces the observations of Cygnus X-1. On the other hand, our model predicts a high energy emission, above 1 TeV, that might be detectable by instruments like MAGIC II or CTA in the near future.

\section{Acknowledgments}

This research was supported by ANPCyT through grant PICT-2007-00848 BID 1728/OC-AR and by the Ministerio de Educaci\'on y Ciencia (Spain) under grant AYA 2007-68034-C03-01, FEDER funds.


\begin{thebibliography}{99}

\bibitem{Narayan} R. Narayan \& I. Yi, \textsl{ApJ} \textbf{428} (1994), L13.

\bibitem{Poutanen} J. Poutanen, J.H. Krolik \& F. Ryde, \textsl{MNRAS} \textbf{192} (1997), L21-L25.

\bibitem{Drury} L. Drury, \textsl{Reports on Progress in Physics} \textbf{46} (1983), 973

\bibitem{Begelman} M. Begelman, B. Rudak \& M. Sikora, \textsl{ApJ} \textbf{362} (1990), 38-51.

\bibitem{Vila} G.S. Vila \& F.A. Aharonian, G.E. Romero \& P. Benaglia (eds), in \emph{Compact Objects and their Emission} (AAA Book Series, Paideia, La Plata, 2009), p. 1-38

\bibitem{McConnell} M.L. McConnell et al., \textsl{ApJ} \textbf{543} (2000), 928-937

\bibitem{Cadolle} M. Cadolle Bel et al., \textsl{A\&A} \textbf{446} (2006), 591

\end{thebibliography}
\end{document}